\documentclass[aps,prb,twocolumn,superscriptaddress,showpacs]{revtex4-1}

\usepackage{graphicx}
\usepackage{calc}
\usepackage{bm}
\usepackage{color}

\begin{document}

\title{Strong coupling between a permalloy ferromagnetic contact and a helical edge channel in a narrow HgTe quantum well}

\author{A.~Kononov}
\affiliation{Institute of Solid State Physics RAS, 142432 Chernogolovka, Russia}
\author{S.V.~Egorov}
\affiliation{Institute of Solid State Physics RAS, 142432 Chernogolovka, Russia}
\author{Z.D. Kvon}
\affiliation{Institute of Semiconductor Physics, Novosibirsk 630090, Russia}
\affiliation{Novosibirsk State University, Novosibirsk 630090, Russia}
\author{N.N. Mikhailov}
\affiliation{Institute of Semiconductor Physics, Novosibirsk 630090, Russia}
\author{S.A. Dvoretsky}
\affiliation{Institute of Semiconductor Physics, Novosibirsk 630090, Russia}
\author{E.V.~Deviatov}
\affiliation{Institute of Solid State Physics RAS, 142432 Chernogolovka, Russia}

\date{\today}

\begin{abstract}
We experimentally investigate spin-polarized electron transport between a permalloy ferromagnet and the edge of a two-dimensional electron system with band inversion, realized in a narrow, 8~nm wide HgTe quantum well. In zero magnetic field,  we observe  strong asymmetry of the edge potential distribution with respect to the ferromagnetic ground lead. This result indicates, that the helical edge channel, specific for the structures with band inversion even at the conductive bulk, is  strongly coupled to  the ferromagnetic side contact, possibly due to the effects of proximity magnetization. It allows selective and spin-sensitive contacting of helical edge states.
\end{abstract}


\maketitle

\section{Introduction}

Recently, there is a strong interest in two-dimensional semiconductor systems with an inverted band structure, like narrow HgTe quantum wells. This interest is mostly connected with the quantum spin-Hall (QSH) effect  regime~\cite{konig,kvon}. Similarly to the conventional quantum Hall (QH) effect in high magnetic fields~\cite{buttiker}, QSH regime  is characterized~\cite{molenkamp_nonlocal,kvon_nonlocal} by presence of two spin-resolved, current-carrying helical edge states~\cite{pankratov,zhang1,kane,zhang2} even in zero magnetic field. The helical QSH edge states are regarded to be suitable for different applications like quantum computing and cryptography. 

Experimental investigations of helical QSH  edge states are mostly based on charge transport along the edge, which has been detected in local and  non-local resistance measurements~\cite{konig,kvon,molenkamp_nonlocal,kvon_nonlocal} and by a direct visualization technique~\cite{imaging}. In the last case, the edge current has even been demonstrated to coexist with the conductive bulk~\cite{imaging}, which is also  possible from theoretical considerations~\cite{pankratov,volkov}. Despite the initial  idea of a topological protection~\cite{konig,zhang1,kane,zhang2}, backscattering appears at macroscopic distances~\cite{kvon,kvon_nonlocal}, possibly due to the allowed two-particle process~\cite{mirlin} or to the electron puddles~\cite{glazman}.

It is clear that for possible applications, it is necessary to develop a  technique of selective contacting of these edge states.  A possible variant is to use spin effects: QSH edge transport is supposed to be essentially spin-dependent~\cite{pankratov,zhang1,kane,zhang2} even in zero magnetic field. Strong coupling between the spin-resolved helical edge states and a ferromagnet can also be anticipated from  theoretical considerations~\cite{qi,lunde}.

Here, we experimentally investigate spin-polarized electron transport between a permalloy ferromagnet and the edge of a two-dimensional electron system with band inversion, realized in a narrow, 8~nm wide HgTe quantum well. In zero magnetic field,  we observe  strong asymmetry of the edge potential distribution with respect to the ferromagnetic ground lead. This result indicates that the helical edge channel, specific for the structures with band inversion even at the conductive bulk, is  strongly coupled to  the ferromagnetic side contact, possibly due to the effects of proximity magnetization. It allows selective and spin-sensitive contacting of helical edge states.

\section{Samples and technique}

\begin{figure}
\includegraphics[width=\columnwidth]{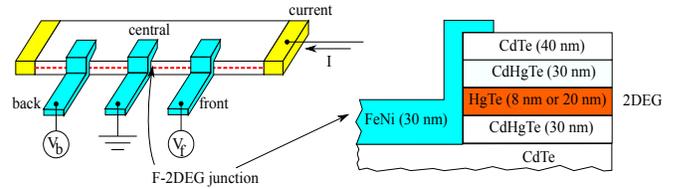}
\caption{(Color online)  Sketch of the type A sample  (not in scale) with electrical connections.  Three ferromagnetic  permalloy $Fe_{20}Ni_{80}$ stripes (denoted as front, central, and back) are placed at the 200~nm mesa step, with low (2-3~$\mu$m) overlap. The width of each stripe is equal to 20~$\mu$m. They  are separated by the 100~$\mu$m distance along the sample edge. In every overlap region, a side junction is formed  between the ferromagnetic  lead and the 2DEG edge.   We  study electron transport through the F-2DEG interface for the central junction in a standard three-point technique: the central ferromagnetic electrode is grounded; a current is applied between it and one of the  normal Au (yellow) contacts; two other ferromagnetic electrodes trace the 2DEG potential to both sides of the grounded junction, $V_f$  and $V_b$, respectively.
}
\label{sample}
\end{figure}

Our $Cd_{0.65}Hg_{0.35}Te/HgTe/Cd_{0.65}Hg_{0.35}Te$ quantum wells are grown by molecular beam epitaxy on GaAs substrate with [013] surface orientations. The layer sequence is shown in Fig.~\ref{sample}, a detailed description can be found elsewhere~\cite{growth1,growth2}. Our wells are characterized by band inversion~\cite{kvon,kvon_nonlocal},  because the wells' width $d=8$~nm is above the critical value 6.3~nm. They contain a 2DEG with the electron  density  of  $1.5 \cdot 10^{11}  $cm$^{-2}$, as obtained from standard magnetoresistance measurements. The 2DEG  mobility at 4K equals to $2\cdot 10^{5}  $cm$^{2}$/Vs. For samples with higher $d=20.5$~nm , a 2D system in the quantum well represents an indirect 2D semimetal~\cite{kvon_s1,kvon_s2}. Both electrons and holes contribute to transport in this case. The carriers' concentrations are low enough, about $0.5 \cdot 10^{11}  $cm$^{-2}$ and $1 \cdot 10^{11}  $cm$^{-2}$ for electrons and holes, respectively.  Electrons' low-temperature mobility is  about $4\cdot 10^{5}  $cm$^{2}$/Vs, because the holes (with lower $5\cdot 10^{4}  $cm$^{2}$/Vs mobility) provide efficient disorder screening~\cite{kvon_scat}.

Our principal idea is to use side contact to the 2DEG edge at the mesa step~\cite{feinas,nbnhgte,nbsemi}. Indeed, the usual procedure with annealed In contacts is not selective with respect to the edge state transport. Annealed In provides high-quality Ohmic contact primary to the bulk 2DEG. Thus, despite the edge current is allowed~\cite{pankratov,volkov} to coexist with the conductive bulk~\cite{imaging}, edge state transport can only be investigated near the charge-neutrality point~\cite{konig,kvon,molenkamp_nonlocal,kvon_nonlocal}. In contrast, without annealing procedure,  the side contact is coupled to the 2DEG edge at the mesa step, because the CdTe layer on the top of the structure is a high-quality insulator at low temperatures.

A sample sketch is presented in Fig.~\ref{sample}. The 100~$\mu$m wide mesa is formed by dry etching (200 nm deep) in Ar plasma.  We fabricate F-2DEG junctions by using rf sputtering to deposit 50~nm thick ferromagnetic permalloy $Fe_{20}Ni_{80}$   stripes at the mesa step, with low (2-3~$\mu$m) overlap.  The stripes are formed by lift-off technique, and the surface is mildly cleaned by Ar plasma before sputtering.  To avoid any 2DEG  degradation, the sample is not  heated during the sputtering process. The source-drain contacts (yellow in Fig.~\ref{sample})  are obtained by thermal evaporation of 100~nm thick Au, as well as the normal Au-2DEG side junctions for reference samples. 

\begin{figure}
\includegraphics[width=\columnwidth]{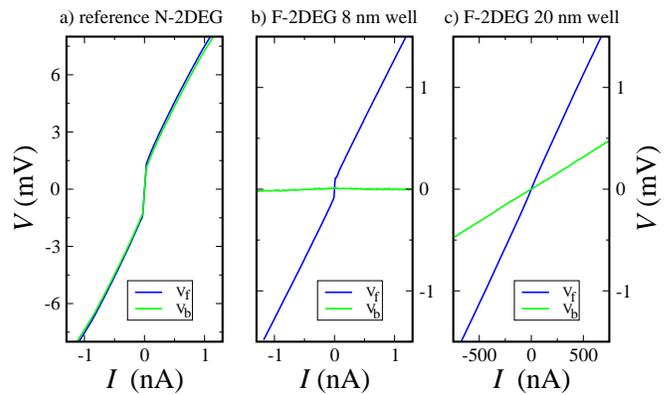}
\caption{(Color online) Examples of  $I-V$ characteristics for transport across a single normal N-2DEG (a) or ferromagnetic F-2DEG (b,c) junction. The measurements are performed at a temperature of 30~mK in zero magnetic field. For  a reference sample with Au-2DEG junctions (a), $V_f$ and $V_b$ coincide well and reflect the resistance of the N-2DEG interface in a standard three-point configuration. For the ferromagnetic F-2DEG junction to the 8~nm HgTe quantum well (b), we obtain significant (about 1~M$\Omega$ corresponding resistance) signal $V_f$, but $V_b$ is always zero.  For the 20~nm HgTe quantum well (c), $V_f$ and $V_b$ are different, however, they are of the same order of magnitude:  we do not observe $V_b=0$  in this case. The data for the 8~nm HgTe quantum well (b) indicate perfect coupling of the the grounded ferromagnetic electrode to the conductive edge channel. 
}
\label{IVnl}
\end{figure}

We study electron transport across one particular F-2DEG junction in a three-point technique, see Fig.~\ref{sample}: the central ferromagnetic electrode is grounded; a current is applied between it and one of the  normal (source or drain) contacts; two other permalloy contacts (front and back) trace the 2DEG potential to both sides  of the grounded junction, $V_f$  and $V_b$, respectively. 

\begin{figure}
\includegraphics[width=\columnwidth]{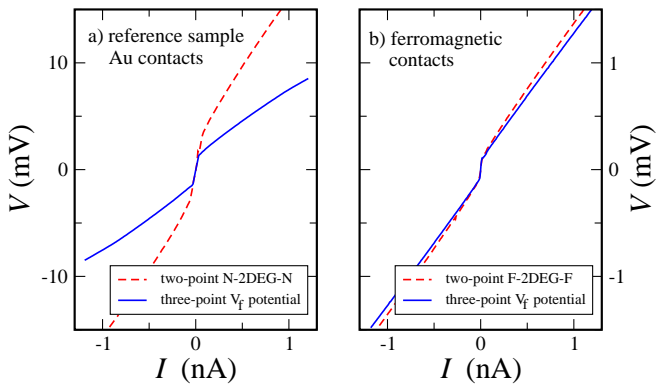}
\caption{(Color online) Two-point $I-V$ characteristics (dash) for double normal   Au-2DEG-Au (a) or  ferromagnetic  F-2DEG-F (b) junctions in comparison with the three-point potential $V_f$ from Fig.~\protect\ref{IVnl} (a) and (b), respectively.   The experimental Au-2DEG-Au $I-V$  reflects the resistance of two mostly identical Au-2DEG interfaces. In contrast, two-point F-2DEG-F  curve  coincides well with the three-point $V_f$ potential, so it reflects the resistance of the edge channel with negligible  interface contributions. The measurements are performed at a temperature of 30~mK in zero magnetic field for the 8~nm wide HgTe well sample.
}
\label{IV2p}
\end{figure}

We sweep the dc current and measure  voltages in a mV range by a dc electrometer, the resulting $I-V$ characteristics are presented in  Fig.~\ref{IVnl},\ref{IV2p}. To obtain $dV/dI(V)$ characteristics in Fig.~\ref{IVgate}, this dc current is additionally modulated by a low ac component (0.01~nA, 2~Hz). We measure the ac ($\sim dV/dI$) component of the 2DEG potential by using a lock-in with a 100~M$\Omega$ input preamplifier. We have checked, that the lock-in signal is independent of the modulation frequency in the range 1~Hz -- 6~Hz, which is defined by applied ac filters. 

The measurements are performed at a temperature of 30~mK. To realize a spin-polarized transport, the permalloy stripes are initially pre-magnetized in the 2DEG plane~\cite{feinas}. The sample is placed within a superconducting solenoid, so the initial in-plane magnetization can be changed by introducing relatively high (above 1~T) external magnetic field. The field is switched to zero afterward, so  the measurements are performed in zero magnetic field.   

Qualitatively similar results were obtained from different samples in several cooling cycles. We study several samples of the type A (from both 8-nm and 20-nm HgTe quantum wells), which are depicted in Fig.~\ref{sample}, and one of the type B, which is additionally covered by a metallic Al gate. The gate covers all the sample (bulk  2DEG, mesa edges, all F-2DEG junctions), except for the normal  Ohmic contacts. To avoid gate leakage, Al gate is  placed over a 350~nm thick dielectric (guanine) layer. We check that there is no noticeable gate leakage through the dielectric at $\pm 5$~V dc gate bias.   As a reference, we use a sample  with Au (normal) side junctions instead of the permalloy ones.

\section{Experimental results}

Examples of  $I-V$ characteristics are presented in Fig.~\ref{IVnl} for transport across a single normal N-2DEG (a) or ferromagnetic F-2DEG (b,c) junction. 

In a three-point technique, the measured potential $V$ reflects in-series connected resistances of the grounded contact and the 2DEG. This technique is especially convenient if the former term is dominant. In this case, the 2DEG is   equipotential, so the measured three-point $I-V$ curve is independent of the particular positions of the current/voltage probes.

This is exactly that we have for the reference normal Au-2DEG junction, see Fig.~\ref{IVnl} (a). $I-V$ curves coincide well for both potential probes $V_f$ and $V_b$. Thus, the measured three-point $I-V$ curves in Fig.~\ref{IVnl} (a) reflect the behavior of the (grounded) Au-2DEG interface. These  $I-V$ curves are obviously non-linear and are characterized by a high resistance (about 10~M$\Omega$) in Fig.~\ref{IVnl} (a). The low-current resistive region with two kinks at about $\pm 1.5$~mV  indicates a significant potential barrier (depletion region~\cite{shklovskii,image02}) at the 2DEG edge. Similar high-resistive junctions we have obtained for other non-magnetic materials like sputtered Nb and NbN~\cite{nbnhgte}.  

Our most prominent experimental result is demonstrated in Fig.~\ref{IVnl} (b). If we ground the permalloy ferromagnetic side contact  to the 8~nm HgTe quantum well, as depicted in Fig.~\ref{sample}, the measured potential is strongly asymmetric. We obtain a significant signal $V_f$, i.e. for the voltage probe placed between the current and ground ones, but $V_b$ is always zero. We observe the same behavior for both current polarities and for two different current probes in Fig.~\ref{sample}, so the asymmetry between $V_f$ and $V_b$ is not connected with any absolute  direction in the sample.  This asymmetry is only determined by  the mutual positions of the current and voltage contacts with respect to the grounded ferromagnetic lead.  Identical behavior is obtained for different ferromagnetic contacts and different 8~nm well  samples. We wish to emphasize that the behavior, depicted in Fig.~\ref{IVnl} (b), is very unusual and is in a high contrast to the standard three-point resistance of a reference Au contact in Fig.~\ref{IVnl} (a). 

The asymmetry $V_f>>V_b=0$  can not originate from bulk 2DEG contribution to the measured potential: different signals $V_f>V_b$ would require the bulk 2DEG resistance to exceed strongly the F-2DEG interface contribution. Because of high-resistive curves  (about 1~M$\Omega$ corresponding resistance) in Fig.~\ref{IVnl} (b), this is  inconsistent with the metallic bulk conductivity (below 1 k$\Omega$) at $1.5 \cdot 10^{11}  $cm$^{-2}$ electron concentration in our samples.

To verify this conclusion experimentally, similar measurements are performed for a 20~nm width HgTe quantum well, see Fig.~\ref{IVnl} (c).  $V_f$ and $V_b$ are also different in this case, however, they are of the same order of magnitude:  we do not observe $V_b=0$  in this case. Both experimental $I-V$ curves correspond to about 1 k$\Omega$ resistance, which is comparable with the bulk values.   In other words, Fig.~\ref{IVnl} (c) experimentally demonstrates the typical effect of bulk current contribution to a three-point signal in the case of low F-2DEG interface resistance. 

The only difference between 8~nm and 20~nm HgTe quantum wells is the conductive helical edge channel in the former case~\cite{molenkamp_nonlocal,kvon_nonlocal}. From the continuous evolution of the edge current when the system is driven away from the charge-neutral regime, demonstrated in Ref.~\onlinecite{imaging} by a direct visualization experiment, one can reasonably suppose that the edge current is still carried by helical spin-resolved edge states even at the conductive bulk 2DEG. It requires  low coupling between the edge states and the bulk, possibly because of the formation of a depletion region where the edge channel is laterally localized~\cite{imaging}. The depletion region of finite width is often  present at the 2DEG edge due to electrostatic effects~\cite{shklovskii,image02}. This depletion region  is also confirmed in our experiment by the zero $V_b=0$ for any distance to the potential probe in Fig.~\ref{IVnl} (b). 

Since the conductive   channel is present at the edge of a 8~nm HgTe quantum well, the perfect ($V_b=0$ for any current $I$) asymmetry  of the edge potential $V_f>>V_b$, observed in Fig.~\ref{IVnl} (b), indicates that the grounded ferromagnetic side electrode is perfectly coupled to this  channel. 

We verify the statement of ideal coupling of the ferromagnetic lead to the edge current by standard two-point  characterization, see Fig.~\ref{IV2p}. It can be seen in the figure, that  the two-point F-2DEG-F $I-V$ curve, measured between two neighbor ferromagnetic contacts, coincides well with the three-point $V_f$ potential from Fig.~\ref{IVnl} (b), so the interface contributions are negligible.
In contrast, the experimental Au-2DEG-Au $I-V$ curve  in Fig.~\ref{IV2p} (a) corresponds to a roughly two times higher resistance than three-point potentials $V_f,V_b$ from Fig.~\ref{IVnl} (a), so it mostly reflects the resistance of two resistive Au-2DEG interfaces.

\section{Discussion} \label{disc}

We should conclude that only a ferromagnetic permalloy side contact is strongly coupled to  the conductive helical edge channel in zero magnetic field:
(i) the perfect ($V_b=0$ for any current $I$) asymmetry  of the edge potential $V_f>>V_b$, see Fig.~\ref{IVnl} (b), is only observed for the 8~nm quantum well; (ii) the coincidence between the two-point F-2DEG-F $I-V$ curve with the three-point $V_f$ potential, see Fig.~\ref{IV2p} (b), directly indicates  negligible F-2DEG interface contributions.  

If the transport current is concentrated at the edge, it is flowing along the shortest edge to the ground lead and there should be no current flowing near the potential probe $V_b$, so $V_b=0$ for any current $I$. The potential $V_f$ in Fig.~\ref{IVnl} (b) reflects therefore solely the resistance of the edge channel between two ferromagnetic contacts. It corresponds to about 1~M$\Omega$ resistance, which is also well known for a transport along the helical current-carrying states at macroscopic distances~\cite{kvon_nonlocal}.  It is worth to mention, that this 1~M$\Omega$ resistance is still much smaller than the resistance for transport through the edge depletion region to the bulk (cp. with Fig.~\ref{IVnl} (a)), so the edge channel is still decoupled from the bulk even at macroscopic distances.

\begin{figure}
\includegraphics[width=0.7\columnwidth]{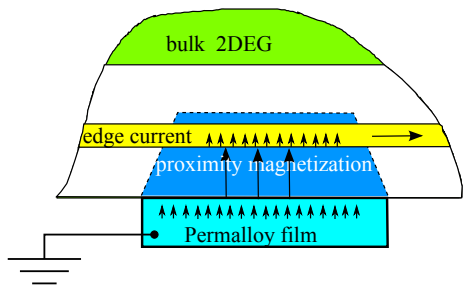}
\caption{(Color online) Top-view of the 2DEG near the ferromagnetic contact. A depletion region (white) is shown, where the edge current is laterally localized (yellow region). A blue region, surrounded by dashed line,  depicts schematically a vicinity of the contact, where the proximity magnetization is important. Here, spin-polarized transport couples edge current with  the ferromagnetic side contact.}
\label{discussion}
\end{figure}

This strong coupling is not defined by chemical composition of the metallic film or the fabrication technique: the sputtered permalloy film contacts the 20~nm HgTe quantum well similarly to other non-magnetic materials. There should be a specific magnetic (spin-dependent) process,  which couples the spin-polarized ferromagnetic side contact and the one-dimensional helical channel at the edge of a 2DEG with band inversion: a proximity magnetization locally aligns~\cite{qi,lunde} the spins of two helical edge states in the vicinity of the ferromagnetic contact.
The spin-polarized electron flow from the ferromagnetic contact can be easily injected to the helical state with corresponding spin projection. The depletion at the interface decouples the bulk 2DEG and makes the helical edge mode even more important.  
Farther transport along the sample edge is diffusive at at macroscopic distances~\cite{kvon,kvon_nonlocal}, because of allowed backscattering~\cite{mirlin,glazman}, so the injected electrons are flowing along the shortest edge to the ground lead. The coupling is independent of the magnetization direction, as we observe in the experiment, because it is the contact magnetization that defines the spin alignment direction, see Fig.~\ref{discussion}.

\begin{figure}
\includegraphics[width=0.9\columnwidth]{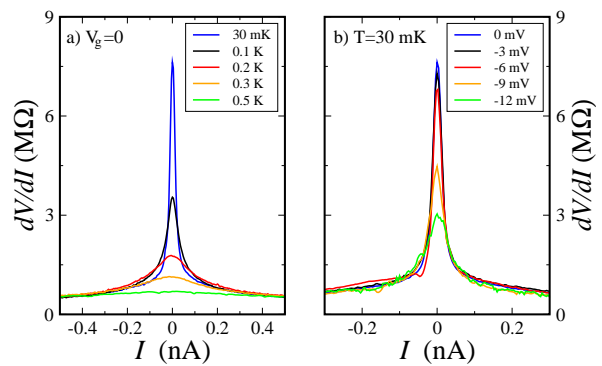}
\caption{(Color online) Differential resistance $dV_f/dI$ at low currents. The  zero-bias resistance peak is strongly affected by (a) temperature and (b) gate voltage. It disappears completely above 0.5~K, which is consistent with the non-linearity onset $\approx 0.06$~mV in Fig.~\protect\ref{IVnl} (b). The measurements are performed in zero magnetic field for the 8~nm wide HgTe well sample B with a metallic gate.
}
\label{IVgate}
\end{figure}

The proximity magnetization  can be directly identified in the experimental data. It opens a gap in the one-dimensional spectrum in the vicinity of the contact~\cite{lunde}, which can be seen as a $\approx 0.06$~mV width region of high resistance in Fig.~\ref{IVnl} (b). This gap only affects the edge channel resistance, and has no effect  on its coupling to the ferromagnetic electrode. 

The zero-bias resistive region is demonstrated in detail in Fig.~\ref{IVgate} as $dV/dI(I)$ dependencies for the sample B with a metallic gate. It disappears completely above 0.5~K, which is consistent in value with the non-linearity onset $\approx 0.06$~mV in Fig.~\ref{IVnl} (b). In contrast, the linear branches of the $dV/dI(I)-I$ curve are invariant below 1~K,  which is also consistent with the reported temperature behavior of the diffusive helical edge state transport at macroscopic distances~\cite{kvon_temp}. The zero-bias resistive region is strongly sensitive to the gate voltage, even if it low enough to  have no effect on the bulk carrier concentration,  see Fig.~\ref{IVgate} (b).  The suppression is fully symmetric with respect to a gate voltage sign. We connect it with the edge state structure reconstruction~\cite{lunde} in the vicinity of a ferromagnetic contact (denoted by a blue region in Fig.~\ref{discussion}), which, however, needs further investigations. A magnetic field above 0.2~T sharply increases the zero-bias resistive region. This behavior is consistent with a spectrum gap~\cite{lunde}  due to proximity magnetization, the gap value can be estimated as   $\approx 0.06$~meV.

\section{Conclusion}
As a result, we experimentally investigate spin-polarized electron transport between a permalloy ferromagnet and the edge of a two-dimensional electron system with band inversion, realized in a narrow, 8~nm wide HgTe quantum well. In zero magnetic field,  we observe  strong asymmetry of the edge potential distribution with respect to the ferromagnetic ground lead. This result indicates that the helical edge channel, specific for the structures with band inversion even at the conductive bulk, is  strongly coupled to  the ferromagnetic side contact, possibly due to the effects of proximity magnetization. It allows selective and spin-sensitive contacting of helical edge states.

\acknowledgments

We wish to thank V.T.~Dolgopolov, V.A.~Volkov, I.~Gornyi, and T.M.~Klapwijk for fruitful discussions.  We gratefully acknowledge financial support by the RFBR, RAS and the Ministry of Education and Science of the Russian Federation under Contract No. 14.B25.31.0007.


\begin{thebibliography}{99}

\bibitem{konig} M. K\"onig, S. Wiedmann, C. Br\"une, A. Roth, H. Buhmann, L. W. Molenkamp, X.-L. Qi, and S.-C. Zhang, Science 318, 766 (2007).
\bibitem{kvon} G. M. Gusev, Z. D. Kvon, O. A. Shegai, N. N. Mikhailov, S. A. Dvoretsky, and J. C. Portal , Phys. Rev. B 84, 121302(R) (2011).
\bibitem{buttiker} M. B\"uttiker, Phys. Rev. B {\bf 38}, 9375 (1988).
\bibitem{molenkamp_nonlocal} A. Roth, C. Br\"une, H. Buhmann, L. W. Molenkamp, J. Maciejko, X.-L. Qi, and S.-C. Zhang, Science 325, 294 (2009).
\bibitem{kvon_nonlocal} G. M. Gusev, E. B. Olshanetsky, Z. D. Kvon, O. E. Raichev, N. N. Mikhailov, and S. A. Dvoretsky, Phys. Rev. B 88, 195305 (2013)

\bibitem{pankratov} B.A. Volkov and O.A. Pankratov, Pis'ma Zh. Eksp. Teor. Fiz. 42, 145 (1985) [JETP Lett. 42, 178 (1985)].
\bibitem{zhang1} S. Murakami, N. Nagaosa, S.-C. Zhang, Phys. Rev. Lett. 93, 156804 (2004).
\bibitem{kane} C. L. Kane, E. J. Mele, Phys. Rev. Lett. 95, 146802 (2005).
\bibitem{zhang2} B. A. Bernevig, S.-C. Zhang, Phys. Rev. Lett. 96, 106802 (2006).

\bibitem{imaging} K.C. Nowack, E.M. Spanton, M. Baenninger, 	M. K\"onig,	J. R. Kirtley,	B. Kalisky,	C. Ames,	P. Leubner,	C. Br\"une,	H. Buhmann,	L. W. Molenkamp,	D. Goldhaber-Gordon	and  K. A. Moler, Nature Materials 12, 787 (2013).

\bibitem{volkov} V.V. Enaldiev, I.V. Zagorodnev, V.A. Volkov, JETP Letters 101,  89 (2015).

\bibitem{mirlin} T.L. Schmidt, S. Rachel, F. von Oppen, and L.I. Glazman, Phys. Rev.  Lett. 108, 156402 (2012); Nikolaos Kainaris, Igor V. Gornyi, Sam T. Carr, and Alexander D. Mirlin, Phys. Rev. B 90, 075118 (2014).
\bibitem{glazman} Jukka I. Vayrynen, Moshe Goldstein, and Leonid I. Glazman, Phys. Rev. Lett. 110, 216402 (2013); Jukka I. Vayrynen, Moshe Goldstein, Yuval Gefen, and Leonid I. Glazman, Phys. Rev. B 90, 115309 (2014)

\bibitem{qi} Xiao-Liang Qi, Taylor L. Hughes, and Shou-Cheng Zhang, Nature Physics 4, 273 (2008)
\bibitem{lunde} Anders Mathias Lunde and Gloria Platero, Phys. Rev. B 86, 035112 (2012).


\bibitem{growth1} Z. D. Kvon, E. B. Olshanetsky, D. A. Kozlov, N. N. Mikhailov, and S. A. Dvoretskii, Pis'ma Zh. Eksp. Teor. Fiz. 87, 588 (2008) [JETP Lett. 87, 502 (2008)].
\bibitem{growth2}E. B. Olshanetsky, Z. D. Kvon, N. N. Mikhailov, E. G. Novik, I. O. Parm, and S. A. Dvoretsky, Solid State Commun. 152, 265 (2012).

\bibitem{kvon_s1} Z. D. Kvon, E. B. Olshanetsky, E. G. Novik, D. A. Kozlov,   N. N. Mikhailov, I. O. Parm, and S. A. Dvoretsky, Phys.
  Rev. B 83, 193304 (2011).
\bibitem{kvon_s2} E. B. Olshanetsky, Z. D. Kvon, Y. A. Gerasimenko, V.   Prudkoglyad, V. Pudalov, N. N. Mikhailov, and S. A.
  Dvoretsky, Pis’ma v ZhETF 98, 947 (2013).
\bibitem{kvon_scat} E. B. Olshanetsky, Z. D. Kvon, M. V. Entin, L. I. Magarill, N. N. Mikhailov, I. O. Parm, and S. A. Dvoretsky, JETP
 Lett. 89, 290 (2009).

\bibitem{feinas} A. Kononov, S.V. Egorov, G. Biasiol, L. Sorba, E.V. Deviatov, Phys. Rev. B 89, 075312 (2014).
\bibitem{nbnhgte} A. Kononov, S. V. Egorov, N. Titova, Z. D. Kvon, N. N. Mikhailov, S. A. Dvoretsky, E. V. Deviatov, 
JETP Letters, 101, 41 (2015).
\bibitem{nbsemi} A. Kononov, S. V. Egorov, Z. D. Kvon, N. N. Mikhailov, S. A. Dvoretsky, and E. V. Deviatov
Phys. Rev. B 93, 041303(R) (2016).

\bibitem{shklovskii} D. B. Chklovskii, B. I. Shklovskii, and L. I. Glazman, Phys. Rev. B {\bf 46}, 4026 (1992).
\bibitem{image02} E. Ahlswede, J. Weis, K. v. Klitzing, K. Eberl, Physica E, {\bf 12}, 165 (2002).

\bibitem{haug} For a review, see, e.g., R. J. Haug, Semicond. Sci. Technol. 8, 131 (1993).







\bibitem{kvon_temp} G. M. Gusev, Z. D. Kvon, E. B. Olshanetsky, A. D. Levin, Y. Krupko, J. C. Portal, N. N. Mikhailov, and S. A. Dvoretsky, Phys. Rev. B 89, 125305 (2014)	



\end{thebibliography}
\end{document}